\documentclass[10pt, a4paper]{article}
\usepackage{lrec2016}
\usepackage{multibib}
\newcites{languageresource}{Language Resources}
\usepackage{graphicx}

\usepackage{epstopdf}
\usepackage[latin1]{inputenc}
\usepackage{hyperref}
\usepackage{csquotes}

\title{Deepfakes in Criminal Investigations:\\ Interdisciplinary Research Directions for CMC Research}

\name{Lorenz Meinen$^{1}$, Astrid Schom\"{a}cker$^{1}$, Stefanie Wiedemann$^{1,2}$, Markus Hartmann$^{3}$, \\ \large\bf Timo Speith$^{1}$, Lena K\"{a}stner$^{1}$, Niklas K\"{u}hl$^{1,2}$, Christian R\"{u}ckert$^{1}$}
\address{$^{1}$ University of Bayreuth, Bayreuth, Germany \\ $^{2}$ Fraunhofer FIT, Bayreuth, Germany, \\ $^{3}$ ZAC NRW, Cologne, Germany\\
         \{firstname.lastname, astrid.schomaecker, lena.kaestner, kuehl, christian.rueckert\}@uni-bayreuth.de\\ 
         }

\abstract{
The emergence of deepfake technologies offers both opportunities and significant challenges. While commonly associated with deception, misinformation, and fraud, deepfakes may also enable novel applications in high-stakes contexts such as criminal investigations. 
However, these applications raise complex technological, ethical, and legal questions. We adopt an interdisciplinary approach, drawing on computer science, philosophy, and law, to examine what it takes to responsibly use deepfakes in criminal investigations and argue that computer-mediated communication (CMC) research, especially based on social media corpora, can provide crucial insights for understanding the potential harms and benefits of deepfakes. Our analysis outlines key research directions for the CMC community and underscores the need for interdisciplinary collaboration in this evolving domain. \\
\newline
 }

\begin{document}

\maketitleabstract

\section{Motivation}

The development and widespread availability of generative AI bears the potential to fundamentally transform human communication. It opens up numerous possibilities, including creating synthetic media using convenient and ready-to-use tools. While some applications of this technology appear unproblematic (e.g., creating animations for school lessons), others pose significant risks. A particularly concerning example are deepfakes: images, videos, or audio files, that convincingly simulate real individuals saying or doing things they never actually did. 

As such, deepfakes are readily associated with deception and fraud, misinformation, or opinion manipulation \cite{Verdoliva}. While this sounds like they are primarily posing threats to society, there is a flip side: Deepfakes may open up possibilities to create useful, realistic media for various contexts. One such context that has yet to receive proper attention is criminal investigations. In criminal investigations, deepfakes could be utilized to infiltrate criminal networks from afar and gain access to privileged information. To that end, investigators would rely on a subtype of deepfakes commonly referred to as (voice or video) \emph{clones}. A clone is the digital impersonation of the voice and/or visage of a particular person, with capabilities of simulating said person in real time, effectively allowing for digital puppeteering. Deploying such clones potentially enables new strategies for infiltration and evidence gathering by replacing undercover agents in certain digital environments. This significantly lowers both risks and costs (in terms of both money and time) usually required to create the relevant identity and gain access to relevant criminal circles. That way, deepfakes might prove vital in the fight against organized crime, which so far has been notoriously challenging to infiltrate.

However, the use of deepfakes in criminal investigation is far from straightforward. Apart from questions of technological feasibility there are significant ethical and legal concerns, particularly regarding the level of deception involved and the safety of the people ``cloned.'' These questions are not only morally but also legally highly relevant, especially since the publication of the AI Act (Regulation (EU) 2024/1689). Yet, scholarly discussions on the potentials and risks of utilizing deepfakes ``for the greater good'' have been sparse.

We believe that the question of how to (legally and morally) utilize deepfakes in criminal investigations needs to be tackled by an interdisciplinary approach. Additionally, in this paper, we suggest several avenues on how it could be further improved with the help of CMC research. We believe that in doing so, our project underscores the potential of analyzing CMC corpora to understand and mitigate the risks associated with deepfakes---a topic which the CMC community has started to actively discuss \cite{russo2024ai}. In the following, we first discuss questions arsing from computer science (\autoref{cs}), philosophy (\autoref{philo}), and legal studies (\autoref{law}), respectively; afterwards, we highlight some important questions that can only be addressed by taking an interdisciplinary perspective (\autoref{inter}). For all these questions, we suggest avenues of how their treatment could be supplemented with CMC research.

\section{Technological Possibilities} \label{cs}
In the context of deepfakes, the primary focus in computer science is on developing and improving increasingly powerful techniques for generating synthetic media and detecting and preventing such fabrications. The ability to create realistic deepfakes has advanced rapidly in recent years \cite{dragar2023beyond}. While early manipulations, so called \enquote{cheapfakes,} were carried out manually or with simple tools and often required significant effort, today's deepfakes are based on complex deep learning architectures, such as Generative Adversarial Networks (GANs), Encoder-Decoder Networks (EDs), Convolutional Neural Networks (CNNs) and Recurrent Neural Networks (RNNs) \cite{Verdoliva,MirskyLeeDeepfakes}. In the context of clones used for criminal investigations, techniques based on these architectures, such as face swapping, face reenactment, and voice conversion, are considered promising but also have considerable potential for misuse.

The technical potential and limitations of these methods warrant careful analysis. Several factors play a role here, including the quality and quantity of training data required to train models that generate realistic deepfakes, data protection issues when using personal data, the availability of suitable hardware and computing resources, and the required expertise in machine learning \cite{Hwang2020AGT}. These technical requirements and barriers vary considerably across user groups. While laypeople or criminal actors often benefit from access to user-friendly software tools, government agencies, such as law enforcement authorities, are subject to strict legal and ethical constraints, for example, regarding data protection and transparency, which can limit their scope of action.

As generative technologies continue to advance, so does research in the field of deepfake detection. Current detection methods typically rely on the analysis of spatial and temporal artifacts, such as blurred object boundaries, inconsistent contextual elements, missing or tampered watermarks, unnatural behavioral patterns, or asynchrony between speech and lip movements \cite{Le}. A thorough understanding of existing detection strategies, along with anticipation of future technical measures to prevent and counteract deepfakes, is crucial for two reasons. First, it allows for reliable identification of forgeries. Second, it ensures that deepfakes used in covert digital operations for criminal investigation purposes cannot easily be unmasked.

\paragraph{Directions for CMC Research.}
To better understand the technical possibilities and limitations of deepfakes, it is important to investigate their current use and creation in real communication environments. CMC researchers could contribute to this effort by answering the following questions: What types of deepfakes are used, and what technologies are used to create them? In this context, it would be useful to empirically investigate which software tools or platforms were used to create these deepfakes, as they may be identifiable based on metadata or file traces. Additionally, CMC research could address the important questions of how easy it is for non-experts to produce convincing deepfakes with publicly available tools and how easily laypeople are convinced by them. Additionally, CMC corpora containing deepfakes could serve as a resource for improving detection systems by evaluating existing methods and providing training material for machine learning-based detection programs.

\section{Moral and Epistemological Implications}\label{philo}
From the philosophical perspective, a central question concerns the ethical permissibility of deepfakes in general, and of clones within criminal investigations specifically. Addressing this question requires weighing potential harms against the potential benefits. 

The philosophical literature highlights several negative consequences associated with deepfakes. Chief among them are deepfakes' capacity to deceive, their potential to erode epistemic trust in media \cite{fallis_epistemic_2021,rini_deepfakes_2020,matthews_deepfakes_2023}, and the violation of the rights of depicted individuals \cite{de_ruiter_distinct_2021,rini_deepfakes_2022}. 
The relationship between deepfakes and our knowledge, i.e. the epistemological effect of deepfakes, is philosophically especially interesting. Starting from the (simplified) definition of knowledge as justified true belief \cite{sep-knowledge-analysis}, deepfakes can affect all aspects of knowledge. 

First, deepfakes inherently convey false information, which can lead to the formation of false beliefs with potentially serious consequences---for example, individuals may fall victim to scams by transferring money to impostors they mistakenly believe to be trusted business partners or loved ones, or voters may be misled into acting against their own interests due to deceptive political content. Second, with an increase in deepfakes, viewers might become more skeptical about the contents of media in general and hence less likely to believe the contents of any recording. Third, and philosophically most complex, the increase in deepfakes might undermine our justification to believe the contents of any audio-visual media. 

The relationship between deepfakes and the rights of the depicted individuals presents a further but no less pressing issue. De Ruiter argues that the distinct moral wrong of deepfakes lies in portraying people in a way to which they would object \cite{de_ruiter_distinct_2021}. Deepfakes can thus be seen to hurt several of a person's moral rights, including privacy, dignity and autonomy. To weigh the severity of the concerns, it can be useful to understand their relationship to the epistemological issues. Worries about the depicted person's rights might be considered less pressing if viewers are unlikely to attribute the contents to the person, either because the fake is easily detectable or because users have generally become skeptical. 

These concerns must be balanced against potential benefits. It has been noted that deepfakes can be used beneficially within creative processes \cite{kerner_beyond_2021}, deepfakes of deceased individuals, so-called deathbots, could have therapeutic use for mourning relatives \cite{lindemann2022ethics} and it has even been hypothesized that deepfakes can \emph{increase} online trust \cite{etienne_future_2021}. Similarly, the use of clones by criminal investigators may have clear positive effects for law enforcement (and thus society at large): it can increase the effectiveness of criminal investigations while simultaneously reducing the risks investigators have to take and the resources that need to be invested. However, if we assume that there are different categories of deepfakes with different moral evaluations then the deepfakes that could be used for criminal investigations are probably among the morally most problematic: They need to resemble a real person convincingly and likely against that person's will, they are meant to deceive the recipient and can put the faked person into harms way. It thus requires careful consideration whether the positive outweighs the negative regarding the use of deepfakes by criminal investigators.

\paragraph{Directions for CMC Research.}
For several of these philosophical questions, it can be useful to investigate how deepfakes are used in CMC and how they are received. One approach can be to investigate the effects of deepfakes on the recipient's beliefs. How likely are internet users to perceive deepfakes as factual images? And how often do they question the veridicality of media shown to them? Can we identify conditions under which recipients are more or less likely to belief in the veridicality of an image? And have these behaviors changed since generative AI has become popularized? Answering these question could help understanding the severity of the effects of deepfakes on our communication and knowledge.
Regarding the rights of depicted persons, corpus analysis could be useful to investigate how closely distributed deepfakes resemble the target person and whether comments on such posts indicate that they can change a viewers opinion of the person. 
Regarding the positive usage, a core question is whether we can identify positive uses of deepfakes in social media or elsewhere, what their numerical relationship is to problematic uses and whether in those cases the positive effects outweigh potential negative ones.

\section{Regulation \& Legal Use of Deepfakes} \label{law}
Taking a look at the EU-law, one finds a vivid discussion on the regulation concerning risks to privacy, issues arising from data protection legislation, such as GDPR and, most recently the AI Act. A key question of these research fields is whether there is a need for more deepfake specific regulation, or if current legislation sufficiently addresses new issues. Even if there was sufficient regulation in theory, one key issue remains the lack of enforceability in the online environment due to anonymity and jurisdiction, rendering existing regulation ineffective \cite{Lantwin2019Deep}. 

A key tool in regulating peoples behavior is criminal law. When focusing on this, one has to narrow their scope, as criminal law is mainly legislated nationally. Therefore (following our expertise) we take a closer look at the German criminal law and criminal proceedings. 
Currently there are efforts to criminalize the publication of deepfakes depicting picture-based sexual violence \cite{Coalitionagreement2025}. This most likely attempts to pass the draft of \S 201b StGB which aims to criminalize the publication of deepfakes violating privacy and intimacy of the depicted (for details see: \cite{BR22224}). This draft addresses concerning trends in computer-mediated communication, however it is not obvious, that there is need for such a law, showcasing the complexity of deepfake specific regulation \cite{Woerlein2024Komplexitaet}.  

The question of deepfake use by criminal investigators has has yet to receive its deserved attention. 
The AI Act assumes the potential legality of clone use by investigators under EU law in Art.~50~(4), by exempting law enforcement from the requirement to disclose the synthetic nature of content for purposes of criminal investigations, if there is a legal basis for said use (for details: \cite{Pehlivan2025} p. 806). Said legal basis is also necessary under national law as every interference with rights requires a legal basis, outlining the extent of interference and the conditions for interference clearly and in a comprehensible manner (in depth: BVerfG, 6. 7. 1999 - 2 BvF 3-90; see \cite{Jarass2024} Rn. 78 f.). Because of the severity of interference with privacy and informational self-determination the investigative powers of \S\S 161, 163 Strafprozessordnung (StPO) do not extend to the creation and deployment of deepfake clones (cf. BVerfG 27.02.2008 - 1 BvR 370/07). Although some parts of creation might have a legal basis within the StPO, under the jurisprudence of the Bundesgerichtshof (BGH), the entire procedure needs to be grounded on a single legal basis (BGH 31.07.2007 - StB 18/06; for in depth analysis: \cite{Rueckert2023digital} p. 469 ff.). With German law lacking the latter, the creation and deployment of clones is not legal in Germany, as of now \cite{Margerie2025Strafverfolgung}. However this is not the only issue concerning the legality of such practices. Both the rights of the person cloned and the person who is misled need to be considered to determine whether a legal basis for the deployment of deepfake clones by law enforcement could even be constitutional. As interference with privacy and informational self-determination weighs in considerably against the legality of said practices, it remains to be seen whether the German legislator undergoes the process of designing a legal basis for the deployment of deepfakes by law enforcement. For creation of a legal basis we will not only need to address national law, but also EU law, as law enforcement gathering personal data and creating a clone falls under Art.~10 Directive (EU) 2016/680 (implemented in \S 48 BDSG). This means the legislator needs to consider the basic rights granted by the German constitution and the Charter of Fundamental Rights of the European Union. In summary, legislation regarding the use of deepfakes in criminal investigation needs to (i) answer the question for which crimes the deployment of clones shall be lawful, (ii) ensure that each individual act of interfering with the rights of the affected (gathering personal data as samples for clone creation, the creation itself, and the deployment of the clone) is addressed, (iii) clarify what safety measures are required when a person is ``cloned,'' and (iv) be in line with the requirements of data protection legislation.

\paragraph{Directions for CMC Research.}
The necessity of deepfake specific criminal regulation in part depends on whether deepfakes are used in a harmful way by the public. Without a good understanding of the status quo of deepfake use and the extent of the harm caused by deepfakes one cannot adequately judge the need for more regulation. Here, CMC research can provide valuable insights by analyzing the use of deepfakes on social media. This analysis would also allow to draw conclusions relevant to understanding the feasibility of deepfake use by investigators, by examining how easily people are deceived by state of the art deepfake technologies. This also gives insights into the requirements for creating convincing deepfakes. 

\section{Interdisciplinary Issues} \label{inter}
Apart from the discipline-specific issues just outlined, there are a number of further inherently interdisciplinary concerns surrounding the use, legitimacy, and implications of the prevalence of deepfakes. These concerns show why our interdisciplinary approach is essential to adequately address the risks and possibilities deepfakes might imply in criminal investigations and beyond. Additionally, answering these research questions could also benefit from insights derived by CMC research.

\paragraph{Trust in Testimony.} 
Linking legal and epistemological considerations, we must wonder how investigators' use of deepfakes within their investigations would square with the use of media as evidence in court. Do we run the risk of jeopardizing the trustworthiness of the legal system if investigators, on the one hand, use deepfakes to deceive suspects and gather evidence, and, on the other hand, provide different recordings as evidence in court? Apart from the general risk of deepfakes undermining trust in media, we need to ask whether the use of deepfakes by specific individuals or outlets undermines their individual perceived trustworthiness. 

CMC research can provide guidance on such questions by analyzing how recipients' behavior changes toward individuals or media after they have been proven to (accidentally or intentionally) spread deepfakes or other fake media. It would be interesting to see whether information about the previous use of fakes leads recipients to be more skeptical in general and question the veridicality of content more than for other outlets. Towards this end, comment sections in social media could be analyzed to look for a change of sentiment.

\paragraph{The Impact of Regulation.}
At the intersection of law and computer science lies the impact of regulation or more precisely the question: How do we design regulation to be impactful and how do we need to implement technological tools to do so? 
As outlined above, the Achilles Heel of regulation in the online environment is often the enforceability, or rather the lack thereof. As good regulation is, among various other factors, characterized by actually impacting peoples behavior, this poses an issue, which needs to be addressed.
Looking at deepfakes specifically, there are two angles new regulation could consider: Regulation could address either the individual or the platform (both are, of course, not mutually exclusive). 

The individual can be addressed by penalizing the publication of certain kinds of deepfakes causing harm, while the platform might be obliged to deny the upload of those deepfakes and/or take down deepfakes under certain conditions. Both angles of regulation need to be supported by technological tools to maximize impact. If we decide to address the individual, we need better ways of overcoming online anonymity. If we decide to address the platforms, we need well-functioning upload filters capable of detecting deepfakes. This means deepfakes need to be disclosed as such in a machine-readable format (as required by Art.~50~(2) AI Act). However, how this obligation is actually fulfilled is a purely technological question. Possible approaches include metadata tagging, digital watermarks or cryptographic provenance systems, whereby the approaches differ in terms of the degree of reliability, manipulability and implementation effort.

Creating meaningful regulation can substantially benefit from empirical insights. CMC research could, for instance, address the following questions: Do people and/or platforms change their behavior when there is new regulation in place which can only be enforced to a limited extend? In which formats are deepfakes actually labeled or disclosed in practice? Following on from this, how do people perceive the different methods of deepfake disclosure and how do they react to them?

\paragraph{Corpora for Criminal Investigations.} 
The issue of how to devise corpora appropriate for criminal investigations lies at the intersection of computer science, philosophy and law. For one thing, as outlined in \autoref{cs}, sufficient quantity and quality of data must be available to train deepfake generation models in order for investigators to successfully use deepfakes---especially clones---to deceive recipients. At the same time, as outlined in \autoref{philo}, there are significant moral and ethical concerns regarding the potential misuse of deepfake technologies and the broader implications of deploying synthetic media in sensitive legal contexts. Finally, as outlined in \autoref{law}, investigators must navigate a complex legal landscape that strictly regulates which types of data may be lawfully collected and used. 

Addressing these challenges requires close interdisciplinary collaboration. Computer scientists must identify what kinds of data are required to meet specific technological objectives; notably, the type of data needed is dictated by the requirements for successful deception, which are only fulfilled when a synthetic artifact achieves perceived authenticity. Ethicists must evaluate the normative implications of using such data and technology in law enforcement. Finally, legal experts must determine whether the collection, processing and storage of the relevant data complies with existing laws and, if not, examine the possibilities of legalizing the collection, processing, and storage.

Against this background, the crucial question becomes how and where relevant data for the creation of clones utilized in criminal investigations can be obtained, processed, and stored in a way that is both legally permissible and ethically sound. It may be interesting to examine whether or to what extent existing (e.g., social media-based) corpora can be legally used in this context. If so, how can we ensure that the data they contain meet the relevant standards?

Answering these questions requires not only technical insight into data adequacy and model performance but also a careful, interdisciplinary examination of the legal and ethical boundaries governing the use of data in criminal investigations as well as expertise in CMC research.

\section{Conclusions \& Outlook}
AI-generated media---deepfakes in particular---bear the potential to change communication significantly. While there are clear risks associated with the increasing dissemination of synthetic media, their widespread availability also offers potentials, e.g., when deepfakes might be utilized ``for the greater good'' to support the rule of law. Yet, even such benevolent uses of deepfakes raise significant ethical and legal questions. To address these, an interdisciplinary perspective is essential. Adding to technological, philosophical, and legal expertise, we believe that systematic corpus-based investigations into how communication is affected by the presence of synthetic media can offer important insights contributing to an effective regulation of deepfakes in criminal investigation and beyond.

\section{Acknowledgments}
Work on this paper has been supported by the project \enquote{For the Greater Good? Deepfakes in Law Enforcement (FoGG)} funded by the Bavarian Institute for Digital Transformation (bidt), an institute of the Bavarian Academy of Sciences and Humanities, under the code KON-024-008.

\section{References}
\bibliographystyle{lrec2016}
\bibliography{xample}

\end{document}